# UV photoprocessing of $CO_2$ ice: a complete quantification of photochemistry and photon-induced desorption processes

R. Martín-Doménech[1], J. Manzano-Santamaría[1], G.M. Muñoz Caro[1], G.A. Cruz-Díaz[1,2], Y.-J. Chen[3,4], V.J. Herrero[5], and I. Tanarro[5]

[1] Centro de Astrobiología (INTA-CSIC), Ctra. de Ajalvir, km 4, Torrejón de Ardoz, 28850 Madrid, Spain e-mail: rmartin@cab.inta-csic.es
[2] Sackler Laboratory for Astrophysics, Leiden Observatory, University of Leiden, PO Box 9513, NL 2300 RA Leiden, the Netherlands
[3] Space Sciences Center and Department of Physics and Astronomy, University of Southern California, Los Angeles, CA90089-1341, USA
[4] Department of Physics, National Central University, Jhongli City, Taoyuan County 32054, Taiwan
[5] Instituto de Estructura de la Materia-Consejo Superior de Investigaciones Científicas (IEM-CSIC), 28006 Madrid, Spain

**ABSTRACT**

*Context.* Ice mantles that formed on top of dust grains are photoprocessed by the secondary ultraviolet (UV) field in cold and dense molecular clouds. UV photons induce photochemistry and desorption of ice molecules. Experimental simulations dedicated to ice analogs under astrophysically relevant conditions are needed to understand these processes.
*Aims.* We present UV-irradiation experiments of a pure $CO_2$ ice analog. Calibration of the quadrupole mass spectrometer (QMS) allowed us to quantify the photodesorption of molecules to the gas phase. This information was added to the data provided by the Fourier transform infrared spectrometer (FTIR) on the solid phase to obtain a complete quantitative study of the UV photoprocessing of an ice analog.
*Methods.* Experimental simulations were performed in an ultra-high vacuum chamber. Ice samples were deposited onto an infrared transparent window at 8K and were subsequently irradiated with a microwave-discharged hydrogen flow lamp. After irradiation, ice samples were warmed up until complete sublimation was attained.
*Results.* Photolysis of $CO_2$ molecules initiates a network of photon-induced chemical reactions leading to the formation of CO, $CO_3$, $O_2$, and $O_3$. During irradiation, photon-induced desorption of CO and, to a lesser extent, $O_2$ and $CO_2$ took place through a process called indirect desorption induced by electronic transitions (DIET), with maximum photodesorption yields ($Y_{pd}$) of ~1.2 x $10^{-2}$ molecules incident photon$^{-1}$, ~9.3 x $10^{-4}$ molecules incident photon$^{-1}$, and ~1.1 x $10^{-4}$ molecules incident photon$^{-1}$, respectively.
*Conclusions.* Calibration of mass spectrometers allows a direct quantification of photodesorption yields instead of the indirect values that were obtained from infrared spectra in most previous works. Supplementary information provided by infrared spectroscopy leads to a complete quantification, and therefore a better understanding, of the processes taking place in UV-irradiated ice mantles.

## 1. Introduction

In the cold and dense regions of the interstellar medium (ISM), icy grain mantles are formed by accretion of gas-phase molecules or formation of molecules on the surface of dust grains. Water is the primary component of these ice mantles. The following most abundant species are CO and $CO_2$, with an abundance of up to ~ 50% relative to water (Mumma & Charnley 2011, and references therein). Molecules in the ice mantles can be energetically processed. In the coldest regions, thermal processing is inhibited, and interaction with photons and cosmic rays plays the most important role. Photoprocessing of ices in molecular clouds is possible thanks to the secondary ultraviolet (UV) field that is generated by the interaction of cosmic rays with $H_2$ molecules that are present in the gas phase (Cecchi-Pestellini & Aiello 1992; Shen et al. 2004). Ice mantles in circumstellar environments can be also irradiated by the central stellar object. UV photons can induce photochemistry and desorption of ice molecules. Photochemistry is initiated by photolysis of absorbing molecules and subsequent recombination of radicals. UV-induced chemistry in ice mantles has been proposed as a possible formation pathway for species of diverse complexity (e.g., Agarwal et al. 1985; Allamandola et al. 1988; Briggs et al. 1992; Bernstein et al. 1995; Gerakines et al. 2001; Moore et al. 2001; Bernstein et al. 2002; Muñoz Caro et al. 2002; Muñoz Caro & Schutte 2003; Gerakines et al. 2004; Muñoz Caro et al. 2004; Loeffler et al. 2005; Meierhenrich et al. 2005; Nuevo et al. 2006; Muñoz Caro & Dartois 2009; Öberg et al. 2009a; Herbst & van Dishoeck 2009; Öberg et al. 2010; Arasa et al. 2013). UV-photon-induced desorption is one of the proposed nonthermal desorption mechanisms, along with grain heating by cosmic rays and exothermic surface reactions (chemical desorption), to explain the presence of molecules in the gas phase at temperatures well below their desorption temperature (e.g., Willacy & Langer 2000; Bergin et al. 2001).

A series of experimental studies have been carried out in the past decades to better understand the photoprocessing of cosmic ice mantles. These studies help understanding astronomical observations of both the solid and gaseous phases of the ISM, and their results are also incorporated into theoretical models of dense clouds and star-forming regions. Despite being the simplest case, experimental simulations on UV photoprocessing of pure ice analogs provide the most fundamental information about the processes triggered by UV photons on ice mantles. They are a necessary step before attempting the simulation of processes in ice mixtures made of various molecular components. Gerakines et al. (1996) presented a study of UV irradiation of nine pure ice analogs of astrophysical interest. Photoprocessing of pure CO ice has been widely studied, since it is one of the most abundant species found in ice mantles. The interface between the UV source and the chamber where the ice sample is located is, usually, a $MgF_2$ window with a cutoff at ∼114 nm (10.87 eV), that is, below the dissociation energy of CO molecules (11.09 eV, Cruz-Díaz et al. 2014a; Chen et al. 2014). Therefore, in these experiments UV irradiation mainly leads to photodesorption (see, e.g., Öberg et al. 2007; Muñoz Caro et al. 2010; Fayolle et al. 2011; Bertin et al. 2012; Chen et al. 2014). In the past years, photoprocessing of other pure ices made of $CO_2$, $H_2O$, $CH_3OH$, $O_2$, or $N_2$ has been studied (see, e.g., Westley et al. 1995; Öberg et al. 2009a,b; Bahr & Baragiola 2012; Fayolle et al. 2013; Yuan & Yates 2013; Fillion et al. 2014; Yuan & Yates 2014; Zhen & Linnartz 2014; Cruz-Díaz et al. 2015).

Ultraviolet photons can induce desorption of ice molecules through two different main mechanisms. Recent works suggest that the so-called photodesorption mainly takes place through a process called indirect desorption induced by electronic transitions (DIET), where UV photons are first absorbed in the subsurface region of the ice, leading to electronic excitation of molecules. Then, the electronic excitation energy is redistributed to the neighboring molecules, providing sufficient energy to the surface molecules to break intermolecular bonds and be ejected into the gas phase (Rakhovskaia et al. 1995; Öberg et al. 2007; Muñoz Caro et al. 2010; Fayolle et al. 2011, 2013; Bertin et al. 2012, 2013; Fillion et al. 2014). Based on molecular dynamics simulations, van Hemert et al. (2015) proposed a direct photodesorption mechanism in which the excited molecule itself desorbs from the ice surface. This mechanism has not yet been confirmed experimentally and may be negligible compared to the indirect photodesorption described above, since the amount of UV photons absorbed in the ice surface is lower than the one absorbed in several subsurface monolayers. However, photodesorption is efficient when the dissociation energy of the absorbing molecules is higher than the photons energy and when these molecules present a high UV-absorption cross section. The other main mechanism for UV-photon-induced desorption is the desorption of photoproducts with enough excess energy immediately after their formation in the surface of the ice. This mechanism is introduced in this paper as photochemical desorption or photochemidesorption, and it includes desorption of photofragments formed with enough kinetic energy and photoproducts resulting from exothermic chemical recombination of radicals (in the surface of the ice in both cases), previously reported in Fayolle et al. (2013) and Fillion et al. (2014). Photochemidesorption is intrinsically different from photodesorption. Certain species do not photodesorb during pure ice irradiation experiments, but photochemidesorption of these species is observed after their formation during irradiation of ices containing parent molecules. Evidence for photochemidesorption is presented in Cruz-Díaz et al. (2015, submitted).

Andersson & van Dishoeck (2008) proposed a third photon-induced desorption mechanism that is also based on molecular dynamics simulations. It is an indirect "kick out" photon-induced desorption mechanism that takes place when an H atom, released from photodissociation of a hydrogen-bearing molecule in the ice, transfers enough momentum to a molecule in the surface of the ice to induce its desorption. However, this mechanism has not yet been confirmed experimentally.

Previous works failed in providing a complete quantitative study of the relative contribution of photodestruction, photoproduction, and photon-induced desorption taking place in irradiated ice analogs, since a quantification of the desorbing molecules in the gas phase is necessary. These molecules are often measured by a quadrupole mass spectrometer (QMS) in addition to the quantification of the solid phase molecules, which usually is performed by infrared spectroscopy. This is not strictly required for the irradiation experiments of pure CO ice, since photodesorption is the main effect (only ∼5% of the absorbed UV photons lead to photoproducts, Muñoz Caro et al. 2010), and can be thus quantified using infrared spectroscopy alone.

In this work, we have used the outcome of our Fourier transform infrared (FTIR) spectrometer and the QMS in previous CO irradiation experiments as a reference for the calibration of the measured QMS ion current of other species. A similar approach has recently been used by Fayolle et al. (2013); Fillion et al. (2014); Zhen & Linnartz (2014) to quantify photon-induced desorption from UV-irradiated pure $O_2$, $N_2$, and $CO_2$ ices, linking the QMS signal of desorbed molecules to the data recorded by reflection absorption infrared spectroscopy and quadrupole mass spectrometry of previous pure CO ice irradiation experiments. Using our calibrated QMS and a transmittance-FTIR spectrometer, we have performed a complete quantitative study of the UV-irradiation experiment of a pure $CO_2$ ice analog. We describe this in Sect. 2. The QMS calibration method for ice irradiation experiments is explained in Appendix A. The experimental results are presented in Sect. 3, while the astrophysical implications are discussed in Sect. 4. The conclusions are summarized in Sect. 5.

## 2. Experimental setup

The UV-irradiation experiments presented in this work have been performed using the InterStellar Astrochemistry Chamber (ISAC) at the Centro de Astrobiología (Muñoz Caro et al. 2010). The ISAC setup is an ultra-high vacuum (UHV) chamber with a base pressure about 4 x $10^{-11}$ mbar, similar to that found in dense cloud interiors. Ice samples were grown by deposition of $CO_2$ (gas, Praxair 99.996%), introduced in the chamber at a pressure of 1 x $10^{-6}$ mbar during 30 s onto a KBr window at 8 K, achieved by means of a closed-cycle helium cryostat. This led to an ice thickness of ∼ 200 x $10^{15}$ $CO_2$ molecules cm$^{-2}$. Additional experiments with $^{13}CO_2$ (gas, Cambridge Isotope Laboratories 99.0%) introduced into the chamber at a pressure of 8.6 x $10^{-6}$ mbar during 1 - 2 min (leading to an ice thickness of ∼ 180 - 320 x $10^{15}$ $CO_2$ molecules cm$^{-2}$) were performed to rule out contamination effects.

Deposited ice analogs were subsequently UV irradiated using an F-type microwave-discharged hydrogen flow lamp (MDHL), from Opthos Instruments, with a $MgF_2$ window used as interface between the lamp and the chamber. This source has a vacuum ultraviolet (VUV) flux of ≈ 2 × $10^{14}$ photons cm$^{-2}$ s$^{-1}$ at the sample position (measured by $CO_2$ → CO actinometry, Muñoz Caro et al. 2010). The spot size of the region irradiated by the lamp coincides with the size of the window substrate. Characterization of the MDHL spectrum has been studied by Chen et al. (2010, 2014), and it is similar to the calculated secondary UV field by Gredel et al. (1989). More information can be found in Cruz-Díaz et al. (2014a), where a description of the VUV spectrophotometer used in our setup to monitor the VUV flux of the hydrogen lamp during ice irradiation is provided.

**Table 1.** IR feature used to calculate the column density of each component (frequencies and band strengths for pure ices at 8 K).

| Molecule | Frequency (cm$^{-1}$) | Band strength (cm molec$^{-1}$) |
|---|---|---|
| $CO_2$ | 2344 | $7.6\times10^{-17}$ [a] |
| $^{13}CO_2$ | 2283 | $7.8\times10^{-17}$ [b] |
| CO | 2141 | $1.1\times10^{-17}$ [c] |
| $^{13}CO$ | 2092 | $1.3\times10^{-17}$ [b] |
| $CO_3$ | 2044 | $1.5\times10^{-17}$ [d] |
| $O_3$ | 1042 | $1.4\times10^{-17}$ [e] |

[a] From Yamada & Person (1964)

[b] From Gerakines et al. (1995)

[c] From Jiang et al. (1975)

[d] This work

[e] From Smith et al. (1985)

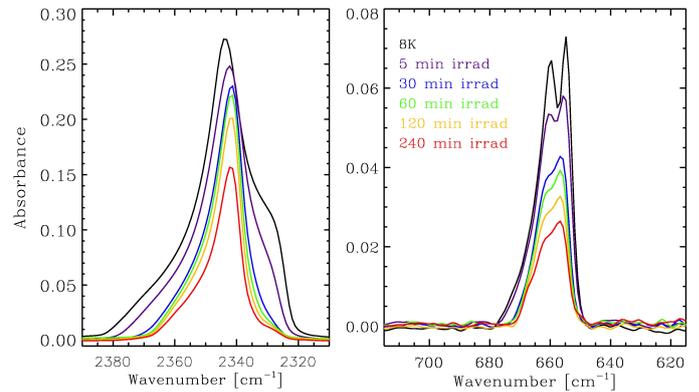

**Fig. 1.** Evolution of two IR bands of $CO_2$ during UV irradiation of a pure $CO_2$ ice analog. *Left*: C=O stretching mode at 2344 cm$^{-1}$. *Right*: degenerate bending modes at ~658 cm$^{-1}$

After irradiation, ice samples were warmed up to room temperature with a heating rate of 1 K min$^{-1}$ using a LakeShore Model 331 temperature controller until complete sublimation was attained, simulating thermal processing of the ice in warm circumstellar regions. A silicon diode temperature sensor was used to measure the sample temperature at any time, reaching a sensitivity of about 0.1 K.

The evolution of the ice after successive irradiations and during the warm-up was monitored by in situ FTIR transmittance spectroscopy, using a Bruker Vertex 70 spectrometer equipped with a deuterated triglycine sulfate detector (DTGS). The IR spectra were collected after ice deposition at 8 K, after every period of UV irradiation, or every five minutes during warm-up, with a spectral resolution of 2 cm$^{-1}$. Column densities of each species in the ice were calculated from the IR spectra using the formula

$$N = \frac{1}{A}\int_{band}\tau_\nu\,d\nu, \quad (1)$$

where $N$ is the column density of the species in molecules cm$^{-2}$, $\tau_\nu$ the optical depth of the absorption band, and $A$ the band strength in cm molecule$^{-1}$, as derived from laboratory experiments (Table 1). The band strength of the 2044 cm$^{-1}$ IR feature corresponding to $CO_3$ was calculated in this work, assuming that all remaining C atoms not locked in $CO_2$ or CO molecules after the first five minutes of irradiation were forming $CO_3$, since no other carbon-bearing photoproducts were detected during the experiments. The other band strengths in Table 1 were measured for pure ices made of one molecular component. The same values are usually adopted in ice mixtures, which introduce an uncertainty of about 20-30% (d'Hendecourt & Allamandola 1986).

Molecules desorbing to the gas phase during irradiation or during the warm-up were detected using a Pfeiffer Prisma quadrupole mass spectrometer (QMS) of mass spectral range from 1 to 200 amu equiped with a Channeltron detector. The QMS ionizes gas-phase molecules with ~ 70 eV electron bombardment. All the species were monitored through its main mass fragment: m/z=44 ($CO_2$), m/z=45 ($^{13}CO_2$), m/z=28 (CO, with a small contribution of $CO_2$ fragmentation into CO$^+$ of, approximately, 10% of the m/z=44 signal), m/z=29 ($^{13}$CO), m/z=60 ($CO_3$), m/z=61 ($^{13}CO_3$), m/z=32 ($O_2$), and m/z=48 ($O_3$).

To convert the signal measured by the QMS into a column density of desorbed molecules during irradiation, we performed additional experiments with the same setup to calibrate our QMS. These experiments are described in Appendix A.

## 3. Experimental results and discussion

### 3.1. IR spectra of $CO_2$ ice during irradiation

As explained in Sect. 2, the solid sample was monitored by IR spectroscopy during the irradiation of the pure $CO_2$ ice. The C=O stretching band of $CO_2$ at 2344 cm$^{-1}$ is the dominant IR feature, although its intensity decreases during UV irradiation of the ice (see left panel of Fig. 1) along with the bands corresponding to combination modes at 3708 cm$^{-1}$ and 3600 cm$^{-1}$ (not shown) and the bending mode at ~658 cm$^{-1}$ (right panel of Fig. 1).

At the same time, three features appear due to the formation of new species upon photochemistry induced by the UV photons. The C=O stretching band of CO at 2141 cm$^{-1}$ rapidly increases after irradiation is initiated, as was reported by Gerakines et al. (1996), reaching its maximum intensity after 60 minutes of irradiation (Fig. 2). A similar behavior is found for the band peaking at 1042 cm$^{-1}$, which corresponds to $O_3$ (Fig. 3). The band peaking at 2044 cm$^{-1}$ is assigned to $CO_3$ (Fig. 2; Gerakines et al. 1996; Öberg et al. 2009b). Unlike CO and $O_3$, this species reaches its maximum abundance during the first five minutes of irradiation, which correspond to a fluence of 6 x 10$^{16}$ photons/cm$^2$, similar to the value found by Öberg et al. (2009b), and then it decreases gradually as the fluence increases. The presence of these photoproducts in the ice leads to the loss of the double-peak structure of the degenerate bending modes of $CO_2$ (see right panel of Fig. 1), as has previously been reported for carbon dioxide in multicomponent ice analogs (Sandford & Allamandola 1990; Martín-Doménech et al. 2014). Similarly, the shoulder of the C=O stretching band of $CO_2$ peaking at ~2328 cm$^{-1}$ corresponding to pure and amorphous $CO_2$ ice gradually decreases during irradiation (Escribano et al. 2013).

### 3.2. Photon-induced desorption during irradiation

The desorbing molecules were detected in the gas phase by the QMS during irradiation. Figure 4 shows the signal of the main mass fragments of all the species mentioned in Sect. 3.1, including m/z = 32, which corresponds to $O_2$. This molecule was not detected in the IR spectra since it is infrared inactive due to the lack of an electric dipole moment. The integrated QMS ion currents have been converted into the column density values shown in Table 3 and Fig. 8 using Eq. A.2, the values in Table 2, and the $k_{CO}$ value found in Sect. A.1.

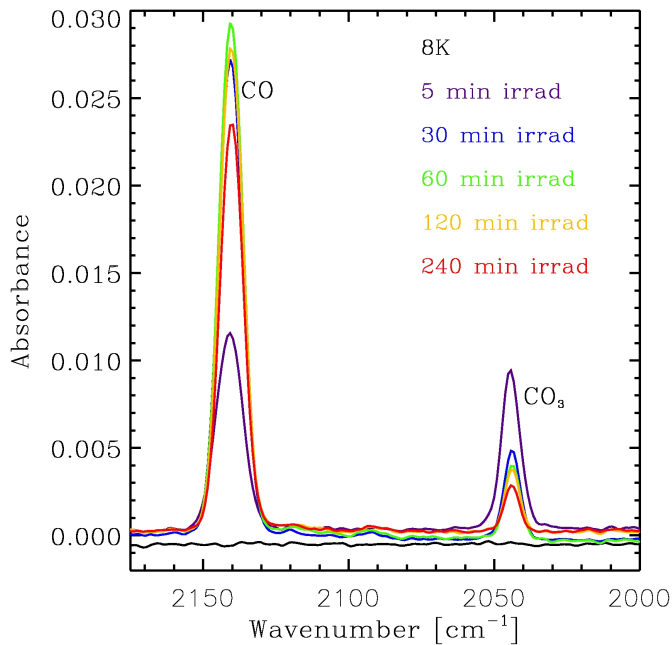

**Fig. 2.** Evolution of the IR bands corresponding to the photoproducts CO and $CO_3$ during UV irradiation of a pure $CO_2$ ice analog.

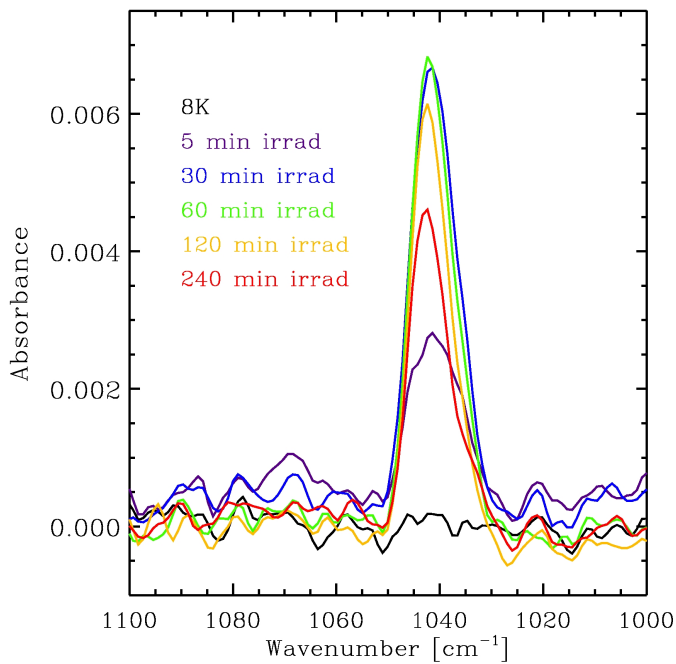

**Fig. 3.** Evolution of the IR feature corresponding to the $O_3$ photoproduct during UV irradiation of pure $CO_2$ ice.

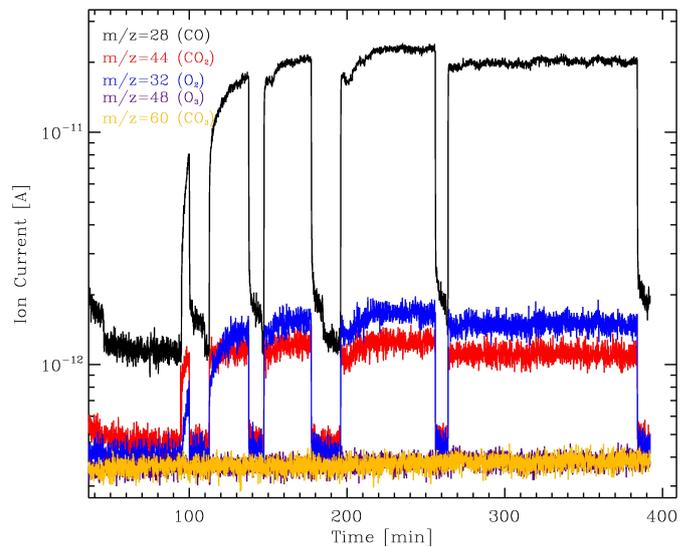

**Fig. 4.** Photodesorption of CO, $O_2$, and $CO_2$ during UV irradiation of pure $CO_2$ ice. The m/z = 44 signal corresponding to $CO_2$ molecules is affected by contamination (see text). Ion current in the y-axis corresponds, approximately, to the partial pressure (mbar) in the main chamber. Note that the y-axis is on a logarithmic scale.

As reported by Bahr & Baragiola (2012); Fillion et al. (2014), the CO product displays the highest desorption in $CO_2$ ice irradiation experiments. The photon-induced desorption of $O_2$ leads to a signal increase over the background level that is about four to five times lower than that of the CO in every irradiation period, which is only slightly lower than the relative increase of the $O_2$ signal observed by Bahr & Baragiola (2012). The increase of the m/z = 44 signal during irradiation is affected by contamination, since a similar rise is detected even when no sample is irradiated in the main chamber (see Fig. 5). This is not the case for $O_2$, which is not detected in Fig. 5. Contamination problems are also reported in Loeffler et al. (2005), with a higher impact since in that work a high vacuum chamber is used instead of a UHV chamber. Irradiation of a $^{13}CO_2$ ice allowed us to rule out the contamination signal and detect the carbon dioxide desorbing from the ice. Figure 6 shows that the amount of $^{13}CO_2$ molecules desorbing during irradiation is even lower than the one corresponding to $O_2$ (see Sect. 3.4). Therefore, photon-induced desorption of $CO_2$ is negligible, as reported in Bahr & Baragiola (2012), in contrast to what was previously claimed in Öberg et al. (2009b). No desorption of $CO_3$ and $O_3$ was detected during irradiation. Photon-induced desorption of these species was not observed in previous works irradiating pure ices (e.g., Bahr & Baragiola 2012; Zhen & Linnartz 2014).

The QMS ion current measured during the various $CO_2$ ice irradiation intervals due to the photon-induced desorption of CO and $O_2$ in Fig. 4 increases progressively with the fluence up to a constant value after approximately 60 minutes of irradiation, which coincides with the irradiation time in which the maximum abundance of CO in the ice is reached. This means that the molecules of both species that formed and accumulated in the bulk of the ice are taking part in the desorption process as the top ice layers are gradually removed by the UV photons. Photon-induced desorption is thus taking place through the DIET mechanism, and photochemidesorption, if taking place, has a negligible effect. If photon-induced desorption were taking place through the photochemical mechanism alone, only molecules newly formed in the surface of the ice would be desorbing. The number of desorbing molecules would then be either constant or, alternatively, it would decrease with continuing irradiation, since there would be fewer $CO_2$ molecules available in the ice surface to lead to photoproducts (see Cruz-Díaz et al. 2015, submitted). Moreover, as explained in Sect. 1, photon-induced desorption of both species has already been observed in pure CO and $O_2$ ice irradiation experiments, respectively. In the case of CO, photochemidesorption does not apply, and photon-induced desorption takes place through the DIET mechanism. Fillion et al. (2014, and references therein) have reported that the contribution of photochemical desorption relative to photodesorption is about ~10%. On the other hand, photodesorption of $CO_2$ was expected to increase with the abundance of CO in the ice according to the experiments performed by Fillion et al. (2014) with a monochromatic UV source. However, evolution of the m/z = 44 and m/z = 45 signals in Figs. 4 and 6 shows that the increase is not significant in our experiments.

**Table 2.** Values used in Eq. A.2 to convert integrated QMS signals to column densities of desorbed molecules.

| Factor | $^{13}CO_2$ | CO | $O_2$ |
|---|---|---|---|
| $\sigma^+(mol)$ (Å$^2$)$^a$ | 3.521$^b$ | 2.516 | 2.441 |
| $I_F(z)$ | 1$^c$ | 1$^c$ | 1$^c$ |
| $F_F(m)^d$ | 0.794$^b$ | 0.949 | 0.898 |
| $k^*_{QMS} \cdot S(m/z)$ (A mbar$^{-1}$ Å$^{-2}$)$^e$ | 4.62 x 10$^{14}$ | 1.03 x 10$^{15}$ | 8.41 x 10$^{14}$ |

$^a$ Extracted from the online database of the National Institute of Standard and Technologies (NIST)

$^b$ We have used the value corresponding to $^{12}CO_2$ as an approximation. This introduces an uncertainty of about 2% in the case of the fragmentation factor, according to previous measurements.

$^c$ A value of 1 has been taken, assuming that no double ionization of the molecules takes place

$^d$ Extracted from the mass spectra library of the QMS software

$^e$ This work (see Sect. A.2)

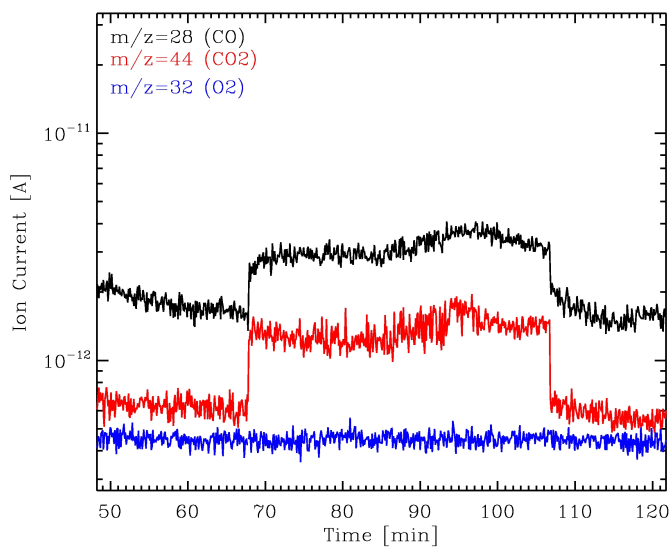

**Fig. 5.** Signals detected by the QMS during irradiation in the main chamber of ISAC without an ice sample. The ion current in the $y$-axis corresponds, approximately, to the partial pressure (mbar) in the main chamber. Note that the $y$-axis is on a logarithmic scale.

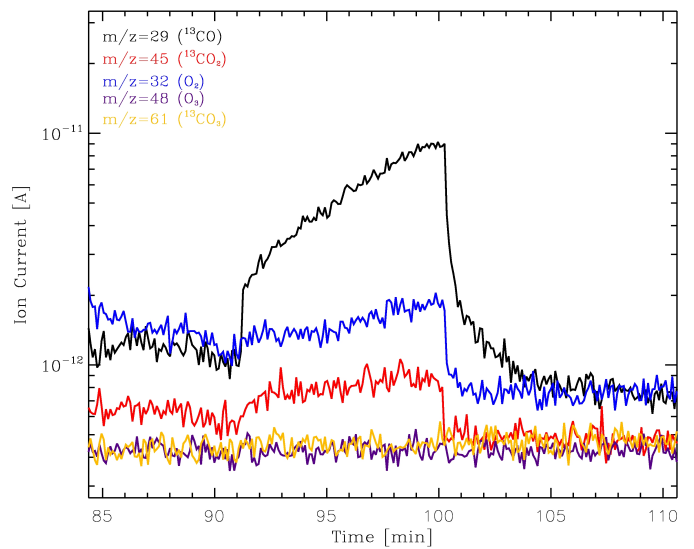

**Fig. 6.** Photodesorption of $^{13}CO$, $O_2$ and $^{13}CO_2$ during UV irradiation of pure $^{13}CO_2$ ice. The ion current in the $y$-axis corresponds, approximately, to the partial pressure (mbar) in the main chamber. Note that the $y$-axis is on a logarithmic scale.

### 3.3. Temperature-programmed desorption of the irradiated $CO_2$ ice

After photoprocessing, the ice sample was warmed up with a heating rate of 1 K/min, leading to thermal desorption of all the observed species. Figure 7 shows the temperature-programmed desorption curves of $CO_2$ and the four photoproducts detected during irradiation.

CO and $O_2$ present desorption peaks at 36 K and 30 K, respectively. While the desorption peak temperature of $O_2$ is similar to that of a pure $O_2$ ice (Acharyya et al. 2007), CO is desorbing at a slightly higher temperature than for pure ice (Muñoz Caro et al. 2010; Martín-Doménech et al. 2014). Both species show a broad peak at ~50 K, which was also detected in Bahr & Baragiola (2012). $O_3$ presents a desorption peak at ~67 K, which is a temperature value similar to that reported in Bahr & Baragiola (2012). Finally, entrapped molecules of the three species co-desorb with $CO_2$ at 87 K, which corresponds to the desorption peak temperature of a pure $CO_2$ ice (see, e.g., Martín-Doménech et al. 2014). Thermal desorption of $CO_3$ is not detected during warm up, probably because of its low abundance at the end of the experiment (see Table 3, leading to a QMS signal below our sensitivity limit, as was the case in Bahr & Baragiola (2012). An additional problem may be the fragmentation pattern of the $CO_3$ ionized molecules, which could not be found in the literature. If m/z=44 (shared with $CO_2$) were the main mass fragment instead of m/z=60 (which is plausible since it is an unstable molecule), detection of $CO_3$ would be more difficult.

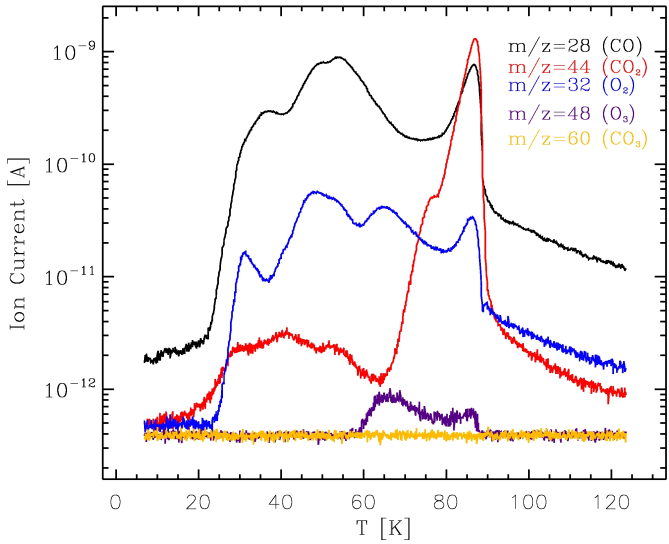

**Fig. 7.** Temperature-programmed desorption curves of $CO_2$ and its photoproducts after UV irradiation of a pure $CO_2$ ice. The ion current in the y-axis corresponds, approximately, to the partial pressure (mbar) in the main chamber. Note that the y-axis is on a logarithmic scale.

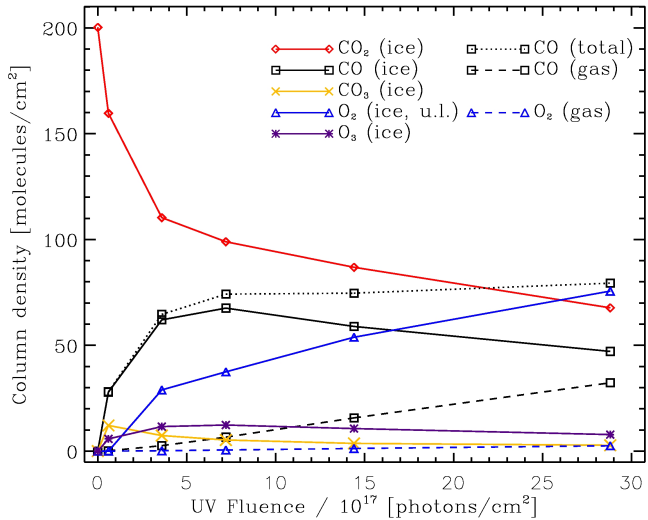

**Fig. 8.** Evolution of the ice and gas column densities of $CO_2$ and its photoproducts during UV irradiation of a pure $CO_2$ ice analog. $O_2$ (ice) column density is an estimated upper limit assuming that all remaining O atoms are locked in $O_2$ molecules.

### 3.4. Photoprocessing of pure $CO_2$ ice: reaction scheme and photodesorption yields

#### 3.4.1. Evolution of the species

Table 3 and Fig. 8 show the evolution of all the species involved in the photoprocessing of a pure $CO_2$ ice analog in both the solid and gaseous phases. Since photon-induced desorption of $CO_2$ is negligible (see Sect. 3.2), we have assumed that $N_{gas}(CO_2) = 0$.

After the first five minutes of irradiation (corresponding to a fluence of 6 x $10^{16}$ photons/cm$^2$), the total number of C and O atoms relative to the initial values in the deposited ice does not remain constant, with a discrepancy up to 25% once the irradiation is completed in the case of C. This is probably because of the uncertainties in the quantification of the column densities in the ice (see Sect. 2), which can be of about 20-30%. In a follow-up paper, the presence of C and O atoms in the ice or in the gas phase after desorption (not quantified here) is studied in more depth, giving an alternative explanation to this issue (Chen et al. 2015). A fraction of O atoms is locked in the ice in the form of $O_2$ molecules, which cannot be detected by IR spectroscopy. We have estimated an upper limit to the column density of $O_2$ in the ice assuming that all remaining O atoms are contained in $O_2$ molecules.

**Table 4.** Initial destruction or formation cross sections of the species during UV irradiation of a pure $CO_2$ ice analog.

| Species | $\sigma_{des/form}$ (cm$^2$)[a] |
|---|---|
| $CO_2$ | $9.5 \times 10^{-18}$ |
| CO | $5.71 \times 10^{-18}$ |
| $CO_3$ | $3.1 \times 10^{-18}$ |
| $O_2$ | $1.1 \times 10^{-18}$ [b] |
| $O_3$ | $1.0 \times 10^{-17}$ |

[a] Uncertainties of 20-30% are expected due to the uncertainty in the measurement of the column densities in the ice.
[b] Upper limit (see text)

An initial destruction or formation cross section $\sigma_{des/form}$ is calculated independently for every species in Table 3, fitting their column densities before equilibrium is reached to an equation of the form

$$N = N_\infty [1 + \frac{N_0 - N_\infty}{N_\infty} e^{-\sigma_{des} Fluence}], \quad (2)$$

or, alternatively,

$$N = N_{max}[1 - e^{-\sigma_{form} Fluence}] \quad (3)$$

for the species that are either destroyed or formed during photoprocessing of the $CO_2$ ice, respectively. These equations are adapted from Loeffler et al. (2005). While $N_0$ and $N_{max}$ are fixed by the experimental data, $N_\infty$ was a free parameter of the fit. Since $CO_3$ is readily formed during the first irradiation period and is then subsequently destroyed, Eq. 2 was used in that case. Calculated cross sections are shown in Table 4. These values should be regarded as apparent initial cross sections for the destruction or formation of every species. The values estimated for $CO_2$ and CO are more than one order of magnitude higher than those reported in Gerakines et al. (1996) for the same experiment, but the ratio between the two is similar. This is probably due to an overestimation of their photon flux, as suggested in Loeffler et al. (2005). At the same time, the value found for the destruction cross section of $CO_2$ is on the same order as the one reported in Loeffler et al. (2005) for the destruction of this molecule in a pure CO ice irradiation experiment, calculated with a more complex model that includes the simultaneous fit of the CO and $CO_2$ column densities at high fluences. A deeper study of the reaction rates would require the assigment of a kinetic constant $k_i$ to every reaction in Sect. 3.4.2 and the simultaneous fit of the column densities of the five species, which is beyond the scope of this work.

**Table 3.** Evolution of the ice and gas column densities of $CO_2$ and its photoproducts during UV irradiation of a pure $CO_2$ ice analog.

| Fluence (photons cm$^{-2}$) | $N_{ice}(CO_2)^a$ | $N_{ice}(CO)$ | $N_{gas}(CO)$ | $N_{total}(CO)$ ($10^{15}$ molecules cm$^{-2}$) | $N_{ice}(CO_3)$ | $N_{ice}(O_2)^b$ | $N_{gas}(O_2)$ | $N_{ice}(O_3)$ |
|---|---|---|---|---|---|---|---|---|
| 0.00 | 200.25 | 0.00 | 0.00 | 0.00 | 0.00 | 0.00 | 0.00 | 0.00 |
| 6.0 x 10$^{16}$ | 159.7 | 27.9 | 0.2 | 28.1 | 12.1 | 0.0 | 0.0 | 5.8 |
| 3.6 x 10$^{17}$ | 110.3 | 62.0 | 2.6 | 64.6 | 7.4 | 28.9 | 0.2 | 11.6 |
| 7.2 x 10$^{17}$ | 99.0 | 67.6 | 6.62 | 74.2 | 5.3 | 37.5 | 0.5 | 12.3 |
| 1.4 x 10$^{18}$ | 87.9 | 58.9 | 15.7 | 74.7 | 3.7 | 53.8 | 1.3 | 10.7 |
| 2.9 x 10$^{18}$ | 67.7 | 47.1 | 32.3 | 79.4 | 2.8 | 75.6 | 2.5 | 7.8 |

$^a$ We have assume that $N_{gas}(CO_2) = 0$ as a first approximation (see text).

$^b$ Estimated upper limit (see text)

### 3.4.2. Reaction scheme

Ultraviolet photons dissociate $CO_2$ molecules through the reaction

$$CO_2 + h\nu \rightarrow CO + O. \quad (4)$$

Carbon monoxide is thus the primary product of photoprocessing of pure $CO_2$ ice. Its final total abundance after a UV fluence of 2.9 x 10$^{18}$ photons/cm$^2$ is ~40% of the initial $CO_2$ column density. This value is four times higher than the one reported in Öberg et al. (2009b) for the photoprocessing of a thinner $CO_2$ ice and a fluence about four times lower, and it is slightly higher than that found in Gerakines et al. (1996) with a similar fluence but a higher initial column density. In our case, with an initial $CO_2$ column density of 200.3 ML, 12.6% of the incident photons are absorbed in the ice (for an average VUV-absorption cross section of 0.67 x 10$^{-18}$ cm$^{-2}$, Cruz-Díaz et al. 2014b), leading to ~0.000125 absorbed photons molecule$^{-1}$ s$^{-1}$.

O atoms formed by the reaction in Eq. 4 can easily react with nearby $CO_2$ molecules to produce $CO_3$ as a secondary product,

$$O + CO_2 \rightarrow CO_3. \quad (5)$$

However, these molecules can be photodissociated, thus re-forming $CO_2$ and O atoms:

$$CO_3 + h\nu \rightarrow CO_2 + O. \quad (6)$$

As the fluence increases, photodissociation dominates formation, and the $CO_3$ abundance drops to ~1% of the initial $CO_2$. This value is lower than the one in Gerakines et al. (1996), but agrees well with the low abundances in more recent works performed in UHV conditions (Öberg et al. 2009b; Bahr & Baragiola 2012).

Alternatively, O atoms can react with the photodissociation products of other $CO_2$ molecules,

$$O + CO + CO_2 \rightarrow 2CO_2 \quad (7)$$

(Bahr & Baragiola 2012)

$$O + O \rightarrow O_2. \quad (8)$$

As irradiation proceeds, reactions 6 and 7 compete with reaction 4, decreasing the net $CO_2$ photodissociation rate to ~10 ML/hr.

In an analogous way, reaction 7 competes with CO formation in reaction 4. Equilibrium is reached after a UV fluence of 7.2 x 10$^{17}$ photons/cm$^2$, at this stage the total number of CO molecules remains almost constant. However, after equilibrium is reached, the column density of CO in the ice decreases due to the contribution of photodesorption, which proceeds at a constant rate of ~8 ML/hr. Finally, ~40% of the produced CO has photodesorbed to the gas phase.

The $O_2$ molecules formed in the ice by reaction 8 can either photodissociate,

$$O_2 + h\nu \rightarrow O + O \quad (9)$$

($E_{dis}$ = 5.1eV for gas phase molecules, Okabe 1978), or further react with free O atoms to form $O_3$,

$$O + O_2 \rightarrow O_3. \quad (10)$$

Ozone molecules can be destroyed by the photodissociation reaction,

$$O_3 + h\nu \rightarrow O_2 + O, \quad (11)$$

leading to a drop in its column density after about one hour of irradiation. Its final abundance is ~4% of the initial $CO_2$ column density, which is slightly lower than the value reported in Gerakines et al. (1996).

### 3.4.3. Photodesorption yields

Photodesorption yields ($Y_{pd}$) of CO and $O_2$ reach constant values after ~ 60 minutes of irradiation (see Table 5), with $Y_{pd}$(CO) ~1.2 x 10$^{-2}$ molecules incident photon$^{-1}$, and $Y_{pd}$($O_2$) ~9.3 x 10$^{-4}$ molecules incident photon$^{-1}$. The value found in this work for $Y_{pd}$(CO), measured directly from the quantification of the CO molecules detected by the QMS is about ten times higher than the one reported in Öberg et al. (2009b) for the sum of all the photodesorption products, which was measured indirectly from the loss of ice molecules using reflection absorption infrared spectroscopy (RAIRS). This photodesorption yield is about five times lower than the one reported in Muñoz Caro et al. (2010); Chen et al. (2014) for pure CO ice irradiation experiments. The photodesorption yield of $O_2$ in this work is ~50% higher than the one reported in Zhen & Linnartz (2014) for VUV irradiation of a pure $O_2$ ice with broad Ly$\alpha$ light. Photodesorption yield of $CO_2$ has only been measured for the $^{13}CO_2$ ice irradiation experiments and remains constant with fluence. We have found a value of $Y_{pd}$ ($^{13}CO_2$) ~1.1 x 10$^{-4}$ molecules incident photon$^{-1}$, on the same order of magnitude as the photodesorption yields reported in Fillion et al. (2014) for monochromatic UV irradiation with similar photon energy. Photodesorption yield of $^{13}$CO is virtually the same as in the case of $^{12}$CO, and we expect a similar behavior for carbon dioxide.

**Table 5.** Evolution of the CO and $O_2$ photodesorption yields during UV irradiation of a pure $CO_2$ ice analog.

| Irradiation time (min) | Fluence (photons cm$^{-2}$) | $Y_{pd}$ (CO)$^a$ $10^{-2}$ ($\frac{\text{molecules}}{\text{incident photon}}$) | $Y_{pd}$ (O$_2$)$^a$ $10^{-4}$ ($\frac{\text{molecules}}{\text{incident photon}}$) |
|---|---|---|---|
| 5 | 6.0 x 10$^{16}$ | 0.3 | 1.7 |
| 30 | 3.6 x 10$^{17}$ | 0.8 | 6.7 |
| 60 | 7.2 x 10$^{17}$ | 1.1 | 9.2 |
| 120 | 1.4 x 10$^{18}$ | 1.3 | 10.0 |
| 240 | 2.9 x 10$^{18}$ | 1.2 | 8.9 |

$^a$ Averaged for each irradiation period. We assume uncertainties of about 30-40%, mostly due to the uncertainty in the ionization cross sections (15% according to Majeed & Strickland 1997) and in the measurement of the UV flux.

## 4. Astrophysical implications

As mentioned in Sect. 1, $CO_2$ is, along with CO, one of the most abundant species found in the water-rich ice mantles covering dust grains. Solid $CO_2$ is consistently detected in a variety of environments. Its average abundance relative to water is ~18% in quiescent molecular clouds where ice mantles are formed (Whittet et al. 2007), and ~17% in high-mass protostellar envelopes (Gerakines et al. 1999). However, this abundance is much higher in circumstellar ices around low-mass protostars (32%, Pontoppidan et al. 2008), suggesting that there is a new formation route to $CO_2$ that is activated in these regions. Formation of solid $CO_2$ is not fully understood yet, but a gas-phase pathway is ruled out since models predict $CO_2$ abundances much lower than observed (Bergin et al. 1995). The two most likely formation pathways are UV photoprocessing of $H_2O$:CO ices (d'Hendecourt et al. 1986) and cosmic-ray irradiation of pure CO ices (e.g., Jamieson et al. 2006). Approximately two thirds of the observed solid $CO_2$ is found in a water-rich environment, while one third is detected in a CO environment. Pure and crystalline $CO_2$ ice, such as the one used in our experimental simulations (Escribano et al. 2013), is also observed. It may be formed by segregation from the water-rich environment or through evaporation of the CO from the CO:$CO_2$ component (Pontoppidan et al. 2008). $CO_2$ ice in space can be energetically processed. In particular, UV photoprocessing in dense cloud interiors is possible thanks to the secondary UV field mentioned in Sect. 1. Assuming a flux value of 10$^4$ photons cm$^{-2}$ s$^{-1}$, ice mantles experience a fluence of 3.2 x 10$^{17}$ photons cm$^{-2}$ in the expected lifetime of a molecular cloud (Shen et al. 2004). This fluence is similar to that experienced for the pure $CO_2$ ice after 30 minutes of irradiation in our experiments.

During our experimental simulations, the estimated photon-induced desorption rate of $CO_2$ was ~1.1 x 10$^{-4}$ molecules per incident VUV photon, which only accounts for the desorption of 3.5 x 10$^{13}$ molecules cm$^{-2}$ during the cloud lifetime. This agrees well with observations, since $CO_2$ is completely frozen, except in very hot or shocked regions (Bergin et al. 1995; Boonman et al. 2003; Nomura & Millar 2004; Lahuis et al. 2007).

Only 4% of the CO formed from photolysis of $CO_2$ molecules photodesorbed after a fluence of 3.6 x 10$^{17}$ photons cm$^{-2}$, although this value increased as irradiation took place in our experiments, up to 75% for a total fluence of 2.88 x 10$^{18}$ photons cm$^{-2}$. Therefore, we do not expect that CO formed from $CO_2$ represents a significant contribution to the gas-phase CO observed in molecular clouds. Instead, photodesorption of pure CO ice has been found to be more efficient by a factor of 5 (Muñoz Caro et al. 2010; Chen et al. 2014).

The column density of $CO_3$ after 30 minutes of irradiation represented only 3.6% of the initial $CO_2$. Moreover, we did not observe photon-induced or thermal desorption of this species during our experiments. The abundance of $CO_3$ in the ISM is thus expected to be very low; this species was indeed never observed. Finally, photodesorption of $O_2$ was very low in our experimental simulations, a rate of 9.3 x 10$^{-4}$ molecules per incident photon was estimated, while photon-induced desorption of $O_3$ was not observed at all. Therefore, these two species would be more likely observed in warm regions where they can be thermally desorbed from the ice mantles. Molecular oxygen has recently been detected in a warm region of the Orion molecular cloud using Herschel observations (Goldsmith et al. 2011).

## 5. Conclusions

We have revisited the UV photoprocessing of a pure $CO_2$ ice analog. IR spectroscopy in transmittance allowed us to monitor the solid sample during irradiation. Three photoproducts were detected: CO, $CO_3$, and $O_3$, with final column densities of approximately 40%, 1%, and 4% relative to the initial $CO_2$ after a total UV fluence of 2.88 x 10$^{18}$ photons cm$^{-2}$. If we limit UV irradiation to a fluence similar to that experienced by the ice mantles in dense cloud interiors (~3 x 10$^{17}$ photons cm$^{-2}$ for a flux of 2 x 10$^{14}$ photons cm$^{-2}$ s$^{-1}$ after a cloud lifetime of 10$^6$ yr, Shen et al. 2004, and references therein), the final column density of CO is slightly lower (32%), while those of $CO_3$ and $O_3$ are higher (4% and 6%, respectively). A fourth photoproduct, the homonuclear $O_2$ with no permanent dipole moment, is IR inactive and could not be detected by the FTIR spectrometer. We have estimated an upper limit to its ice abundance of 38% relative to the initial $CO_2$ for a fluence of 2.88 x 10$^{18}$ photons cm$^{-2}$, or 14% after a fluence similar to the above astrophysical scenario. A chemical network leading to these four photoproducts was presented in Sect. 3.4. We used a calibration method for quadrupole mass spectrometers that allowed us to quantify the photodesorbing molecules detected directly in the gas phase during ice irradiation, leading to an estimation of the fraction of photoproducts in the ice that were photodesorbed. Photodesorption of CO and, to a lesser extent, $O_2$ and $CO_2$ was observed in the $CO_2$ ice irradiation experiment, with photodesorption yields of $Y_{pd}$ (CO) ~1.2 x 10$^{-2}$ molecules incident photon$^{-1}$, $Y_{pd}$ ($O_2$) ~9.3 x 10$^{-4}$ molecules incident photon$^{-1}$, and ~1.1 x 10$^{-4}$ molecules incident photon$^{-1}$, respectively. Photon-induced desorption took place mainly through the DIET mechanism, since the QMS signal produced by CO and $O_2$ desorbed molecules gradually increases until the maximum column density of CO is reached as a result of the participation of accumulated molecules in the bulk of the ice. These findings have important implications for the release of ice molecules into the gas phase in cold regions where thermal desorption is inhibited. Certain molecules, like $CO_2$ in this paper, are efficiently photodissociated and their photodesorption is negligible, while photon-induced desorption of their photoproducts is observed, driven either by the DIET mechanism (as in this work) or by photochemidesorption (see Cruz-Díaz et al. 2015, submitted). The photodesorption yields found in this work are not high enough to represent a significant contribution to the presence of these species in the gas phase of quiescent molecular clouds, since only 4% of the formed CO, and a negligible amount of $CO_2$ and $O_2$ have photodesorbed after a UV fluence similar to the one that ice mantles experience during the mean lifetime of these regions. Therefore, the observed gaseous CO in molecular clouds may come from a different origin than $CO_2$ photolysis, while gaseous $CO_2$, $O_2$, and $O_3$ are more likely present in warm regions of circumstellar environments where these species can be thermally desorbed. Finally, based on our experimental evidence, $CO_3$ production may be too low to be observed in the gas phase.

*Acknowledgements.* This research was financed by the Spanish MINECO under project AYA2011-29375. R.M.D. benefited from a FPI grant from Spanish MINECO. V.J.H. and I.T. acknowledge financial support from Spanish MINECO under project FIS2013-48087-C2-1-P.

## References

Acharyya, K., Fuchs, G. W., Fraser, H. J., van Dishoeck, E. F., & Linnartz, H. 2007, A&A, 466, 1005
Agarwal, V. K., Schutte, W. A., Greenberg, J. M., et al. 1985, Origins of Life and evolution of the Biosphere, 16, 21
Allamandola, L. J., Sandford, S. A., & Valero, G. J. 1988, Icarus, 76, 225
Andersson, S., & van Dishoeck, E. F. 2008, A&A, 491, 907
Arasa , van Hermet, M. C., van Dishoeck, E. F., & Kroes, G. J. 2013, JPCA, 117, 7064
Bahr, D. A. & Baragiola, R. A. 2012, ApJ, 761, 36
Bergin, E. A., Ciardi, D. R., Lada, C. J., et al. 2001, ApJ, 557, 209
Bergin, E. A., Langer, W. D., & Goldsmith, P. F. 1995, ApJ, 441, 222
Bernstein, M. P., Sandford, S. A., Allamandola, L. J., Chang, S., & Scharberg, M. A. 1995, ApJ, 454, 327
Bernstein, M. P., Dworkin, J. P., Sandford, S. A., Cooper, G. W., & Allamandola, L. J. 2002, Nature, 416, 401
Bertin, M., Fayolle, E. C., Romanzin, C., et al. 2012, PCCP, 14, 9929
Bertin, M., Fayolle, E. C., Romanzin, C., et al. 2012, ApJ, 779, 120
Boonman, A. M. S., van Dishoeck, E. F., Lahuis, F., & Doty, S. D. 2003, A&A, 399, 1063
Briggs, R., Ertem, G., Ferris, J. P., et al. 1992, Origins of Life and evolution of the Biosphere, 22, 287
Cecchi-Pestellini, C. & Aiello, S. 1992, MNRAS, 258, 125
Chen, Y.-J., Chu, C.-C, Lin, Y.-C. et al. 2010, Advances in Geosciences, 25, 259
Chen, Y.-J., Chuang, K.-J, Muñoz Caro, G. M., et al. 2014, ApJ, 781, 15
Chen, Y.-J., Chuang, K.-J, Muñoz Caro, G. M., et al. in preparation
Cruz-Díaz, G. A., Muñoz Caro, G. M. et al. 2015, A&A, submitted
Cruz-Díaz, G. A., Muñoz Caro, G. M., Chen, Y.-J., & Yih, T.-S. 2014a, A&A, 562, A119
Cruz-Díaz, G. A., Muñoz Caro, G. M., Chen, Y.-J., & Yih, T.-S. 2014b, A&A, 562, A120
d'Hendecourt, L. B.,& Allamandola, L. J. 1986, A&AS, 64, 453
d'Hendecourt, L. B., Allamandola, L. J., Grim, R. J. A., & Greenberg, J. M. 1986, A&A, 158, 119
Escribano, R. M., Muñoz Caro, G. M., Cruz-Díaz, G. A.,, Rodríguez-Lazcano, Y., & Maté, B. 2013, PNAS, 110, A32
Fayolle, E. C., Bertin, M., Romanzin, C., et al. 2011, ApJL, 739, L36
Fayolle, E. C., Bertin, M., Romanzin, C., et al. 2013, A&A, 556, A122
Fillion, J.-H., Fayolle, E., Michaut, X., et al. 2014, Faraday Discuss., 168, 533
Gerakines, P. A., Schutte, W. A., Greenberg, J. M., & van Dishoeck, E. F. 1995, A&A, 296, 810
Gerakines, P. A., Schutte, W. A., & Ehrenfreund, P. 1996, A&A, 312, 289
Gerakines, P. A., Whittet, D. C. B., Ehrenfreund, P. A., et al. 1999, ApJ, 522, 357
Gerakines, P. A., Moore, M. H., & Hudson, R.L. 2001, J. Geophys. Res. -Planets, 106 (E12), 33381
Gerakines, P. A., Moore, M. H., & Hudson, R.L. 2004, Icarus, 170, 202
Goldsmith, P. F., Liseau, R., Bell, T. A., et al. 2011, ApJ, 737, A96
Gredel, R., Lepp, S., Dalgarno, A., & Herbst, E. 1989, ApJ, 347, 289
Herbst, E., & van Dishoeck, E. F. 2009, Annu. Rev. Astron. Astrophys., 47, 427
van Hemert, M. C., Takahashi, J., & van Dishoeck, E. F. 2015, A&A, in preparation
Jamieson, C. S., Mebel, A. M., & Kaiser, R. I. 2006, ApJS, 163, 184
Jiang, G. J., Person, W. B., & Brown, K. G. 1975, J. Chem. Phys., 62, 1201
de Laeter, J. R., Böhlke. J. R., De Bievre, P., et al. 2003, Pure and Applied Chemistry, 75, 685
Lahuis, F., van Dishoeck, E. F., Blake, G. A., et al. 2007, ApJ, 665, 492
Loeffler, M. J., Baratta, G. A., Palumbo, M. E., Strazzulla, G., & Baragiola, R. A. 2005, A&A, 435, 587
Majeed, T. & Strickland, D. J. 1997, J. Phys. Chem. Ref. Data, 26, 2, 335
Martín-Doménech, R., Muñoz Caro, G. M., Bueno, J., & Goesmann, F. A&A, 2014, 564, A8
Meierhenrich, U. J., Muñoz Caro, G. M., Schutte, W. A., et al. 2005, Chem. Eur. J., 11, 4895
Moore, M. H., Hudson, R. L., & Gerakines, P. A. 2001, Spectrochimica Acta Part A, 57, 843
Mumma, M. J., & Charnley, S.B., 2011, Annu. Rev. Astron. Astrophys., 49, 471
Muñoz Caro, G. M., Meierhenrich, U. J., Schutte, W. A., et al. 2002, Nature, 416, 403
Muñoz Caro, G. M., Meierhenrich, U. J., Schutte, W. A., Thiemann, W. H. P., & Greenberg, J. M. 2004, A&A, 413, 209
Muñoz Caro, G. M., Ruiterkamp, R., Schutte, W. A., Greenberg, J. M., & Mennella, V. 2001, A&A, 367, 347
Muñoz Caro, G. M., & Schutte, W. A. 2003, A&A, 412, 121
Muñoz Caro, G. M., & Dartois, E. 2009, A&A, 494, 109
Muñoz Caro, G. M., Jiménez-Escobar, A., Martín-Gago, J.Á. et al. 2010, A&A, 522, A108
Nomura, H., & Millar, T. J. 2004, A&A, 414, 909
Nuevo, M., Meierhenrich, U. J., Muñoz Caro, G. M., et al. 2006, A&A, 457, 741
Öberg, K. I., Fuchs, G. W., Awad, Z., et al. 2007, ApJ, 662, L23
Öberg, K. I., Garrod, R. T., van Dishoeck, E. F., & Linnartz, H. 2009a, A&A, 504, 891
Öberg, K. I., van Dishoeck, E. F., & Linnartz, H. 2009b, A&A, 496, 281
Öberg, K. I., van Dishoeck, E. F., Linnartz, H., & Andersson, S. 2010, ApJ, 718, 832
Okabe, H. 1978. Photochemistry of small molecules, ed. John Wiley & Sons, New York.
Pontoppidan, K. M., Boogert, A. C. A., Fraser, H. J., et al. 2008, ApJ, 678, 1005
Rakhovskaia, O., Wiethoff, P., & Feulner, P. 1995, NIM B, 101, 169
Rejoub, R., Lindsay, B. G., & Stebbings, R. F. 2002, Phys. Rev. A, 65, 042713
Sandford, S. A. & Allamandola, L. J. 1990, ApJ, 355, 357
Shen, C. J., Greenberg, J. M., Schutte, W. A., & van Dishoeck, E. F. 2004, A&A, 415, 203
Smith, M. A. H., Rinsland, C. P., Fridovich, B., & Rao, K. N. 1985, in Molecular Spectroscopy-Modern research, Vol. III, ed. K.N.Rao (Academic Press, London), 111
Tanarro, I., Herrero, V. J., Islyaikin, A. M., et al. 2007, J. Phys. Chem. A, 111, 9003
Westley, M. S., Baragiola, R. A., Johnson, R. E., & Baratta, G. A. 1995, Nature, 373, 405
Willacy, K., & Langer, W. D. 2000, ApJ, 544, 903
Whittet, D. C. B., Shenoy, S. S., Bergin, E. A., et al. 2007, ApJ, 655, 332
Yamada, H., & Person, W. B. 1964, J. Chem. Phys., 41, 2478
Yuan, C. & Yates, Jr., J. T. 2013, J. Chem. Phys., 138, 154303
Yuan, C. & Yates, Jr., J. T. 2014, ApJ, 780, 8
Zhen, J., & Linnartz, H. 2014, MNRAS, 437, 3190

## Appendix A: Calibration of QMS

The integrated ion current measured by a QMS corresponding to a mass fragment $m/z$ of the molecules of a given species desorbed during ice irradiation experiments is proportional to the total number of molecules desorbed, and it can be calculated as follows:

$$A(m/z) = k_{QMS} \cdot \sigma^+(mol) \cdot N(mol) \cdot I_F(z) \cdot F_F(m) \cdot S(m/z), \quad (A.1)$$

where $A(m/z)$ is the integrated area below the QMS signal of a given mass fragment $m/z$ during photon-induced desorption, $k_{QMS}$ is the proportionality constant, $\sigma^+(mol)$ the ionization cross section for the first ionization of the species of interest and the incident electron energy of the mass spectrometer, $N(mol)$ the total number of desorbed molecules in column density units, $I_F(z)$ the ionization factor, that is, the fraction of ionized molecules with charge $z$, $F_F(m)$ the fragmentation factor, that is, the fraction of molecules of the isotopolog of interest leading to a fragment of mass $m$ in the mass spectrometer, and $S(m/z)$ the sensitivity of the QMS to the mass fragment $(m/z)$ (see Tanarro et al. 2007, and references therein).

The measured ion current depends on the ionization cross section of the species ($\sigma^+(mol)$), since only ionized molecules produce a signal in the QMS. The product $I_F(z) \cdot F_F(m)$ represents the fraction of desorbed molecules that leads to the monitored mass fragment $m/z$ once they are ionized in the QMS. The sensitivity $S(m/z)$ must be calibrated for every mass spectrometer. On the other hand, $k_{QMS}$ depends not only on the mass spectrometer, but on the configuration of the experimental setup, and, in particular, on the fraction of desorbed molecules that reach the QMS. This fraction in turn depends on the pumping speed for the different molecules. A brief study of the pumping speed of the setup described in Sect. 2 is presented in Muñoz Caro et al. (2010). In the following discussion we have assumed as a first approximation that this speed is the same for all the molecules, so that $k_{QMS}$ does not depend on the species. The same assumption has been used in previous works (see Fayolle et al. 2013; Fillion et al. 2014, and references therein). Both $k_{QMS}$ and, to a lesser extent, $S(m/z)$ are subject to variations in the behavior of the QMS with time. Calibration of these two factors for our QMS is described below.

*Appendix A.1: Calibration of proportionality constant $k_{QMS}$*

As we explained in Sect. 1, CO ice irradiation experiments under UHV conditions can be used as a reference to calibrate the QMS signal. Equation A.1 can be rewritten as follows:

$$A(m/z) = k_{CO} \cdot \frac{\sigma^+(mol)}{\sigma^+(CO)} \cdot N(mol) \cdot \frac{I_F(z)}{I_F(CO^+)} \cdot \frac{F_F(m)}{F_F(28)} \cdot \frac{S(m/z)}{S(28)}, \quad (A.2)$$

where $k_{CO}$ is

$$k_{CO} = \frac{A(28)}{N(CO)} = k_{QMS} \cdot \sigma^+(CO) \cdot I_F(CO^+) \cdot F_F(28) \cdot S(28). \quad (A.3)$$

Therefore, the problem of calibrating $k_{QMS}$ can be replaced by the problem of calculating $k_{CO}$. To derive $k_{CO}$ for our experimental setup, we performed UV-irradiation experiments of a pure CO ice made by deposition of CO (gas, Praxair 99.998%), using the protocol described in Sect. 2. Since in the case of solid CO photon-induced chemistry accounts for only ~5% of the absorbed photons (Muñoz Caro et al. 2010), the area below the m/z=28 signal detected with the QMS ($A(28)$ in Eq. A.3; left panel of Fig. A.1) practically corresponds to the loss of CO molecules in the ice ($N(CO)$ in Eq. A.3), measured with the FTIR spectrometer (middle panel of Fig. A.1). The relation between $A(28)$ and $N(CO)$ (right panel of Fig. A.1) gives $k_{CO} = 1.32 \times 10^{-10}$ A min ML$^{-1}$, with 1 ML defined as $10^{15}$ molecules cm$^{-2}$. This value is good for our setup by the time the experiments were performed.

**Table A.1.** Values used in Eq. A.5 to derive the relation between $k^*_{QMS} \cdot S$ and $m/z$.

| Factor | He | Ne | Ar |
|---|---|---|---|
| $\sigma^+(mol)$ (Å$^2$)[a] | 0.296 | 0.475 | 2.520 |
| $I_F(z)$[a] | 1 | 1 | 0.948 |
| $Is_F(m)$[b] | 1.000 | 0.905 | 0.996 |

[a] From Rejoub et al. (2002)
[b] From de Laeter et al. (2003)

*Appendix A.2: Calibration of the QMS sensitivity $S(m/z)$*

The experimental protocol used to obtain $S(m/z)$ was slightly different from the one described in Sects. 2 and A.1. The QMS ion current ($I(m/z)$), corresponding to a mass fragment $m/z$ of the molecules of a given species present in the chamber at a given time, is proportional to the pressure measured for that species ($P_{mol}$), and can be calculated with an equation similar to Eq. A.1, but with a different constant of proportionality:

$$I(m/z) = k^*_{QMS} \cdot \sigma^+(mol) \cdot P(mol) \cdot Is_F(m) \cdot I_F(z) \cdot F_F(m) \cdot S(m/z). \quad (A.4)$$

In this case, the Bayard-Alpert ionization gauge (model Ion-iVac IM540) used as manometer in ISAC does not discriminate between different isotopologs of the same species, and $P(mol)$ (once corrected for the gas correction factors provided by the gauge manufacturer) account for all of them. Therefore, an isotopic factor ($Is_F(m)$) must be introduced to take into account only the isotopolog of mass $m$ that is being measured by the QMS.

Since Eq. A.2 uses the relative sensitivity of the QMS between any mass fragment and the mass fragment m/z=28, it is not necessary to obtain the absolute sensitivity of the mass spectrometer. Instead we worked with

$$k^*_{QMS} \cdot S(m/z) = \frac{I(m/z)}{\sigma^+(mol) \cdot P(mol) \cdot Is_F(m) \cdot I_F(z) \cdot F_F(m)} \quad (A.5)$$

for every mass fragment, since the ratio $S(m/z)/S(28)$ is the same as the ratio $k^*_{QMS} \cdot S(m/z)/k^*_{QMS} \cdot S(28)$.

To derive a relation between $k^*_{QMS} \cdot S$ and $m/z$ that could be used for mass fragments of any species, we introduced three nobel gases into the ISAC chamber: He (gas, Air Liquide 99.999%), Ne (gas, Air Liquide 99.995%), and Ar (gas, Praxair 99.997%) at different pressures. For every gas, the ratio between the QMS signal ($I(m/z)$) and the pressure of the gas ($P(mol)$) was corrected for the factors in Eq. A.5 (Table A.1, except for the fragmentation factor, since noble gas atoms do not produce any fragments in the mass spectrometer), leading to a value of $k^*_{QMS} \cdot S(m/z)$ for three different $m/z$ values (m/z=4 for He, m/z=20 for Ne, and m/z=40 for Ar). These values are plotted in Fig. A.2. An exponential fit leads to a sensitivity curve

$$k^*_{QMS} \cdot S(m/z) = 4.16 \times 10^{15} \cdot e^{-(m/z)/20.02} \quad (A.6)$$

in A mbar$^{-1}$ Å$^{-2}$

R. Martín-Doménech et al.: UV photoprocessing of $CO_2$ ice: photochemistry and photon-induced desorption

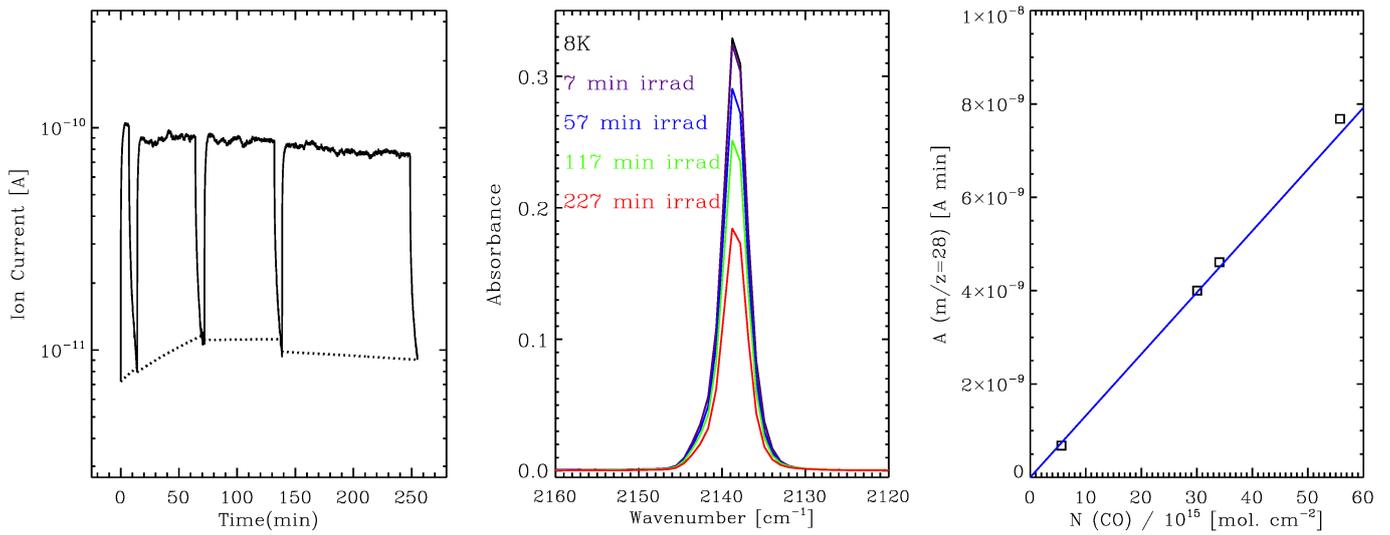

**Fig. A.1.** *Left*: Evolution of the $m/z = 28$ QMS signal corresponding to the photodesorption of CO during UV irradiation of a pure CO ice. The ion current in the *y*-axis corresponds, approximately, to the partial pressure (mbar) in the main chamber. Dotted lines represent the baselines used to calculate $A(28)$ in every irradiation period. Note that the *y*-axis is on a logarithmic scale. *Middle*: Evolution of the C=O stretching mode of CO at 2139 $cm^{-1}$ during UV irradiation of pure CO ice. Differences in the column densities after every period of irradiation correspond to $N(CO)$ in Eq. A.3. *Right*: Relation between $A(28)$ and $N(CO)$. The solid line is a linear fit.

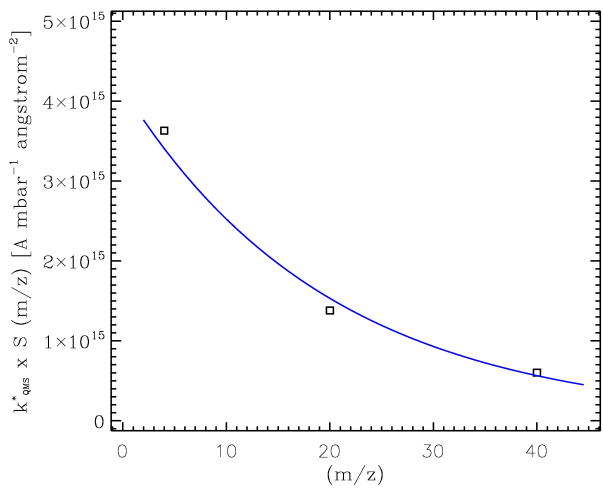

**Fig. A.2.** Relation between $k^*_{QMS} \cdot S(m/z)$ and $m/z$. The solid line is an exponential fit.